\documentclass[slac_one]{revtex4}
\usepackage{graphicx}
\usepackage{amssymb}
\usepackage{abbrevs}

\usepackage{fancyhdr}
\newcommand{\mhmax}{\ensuremath{m_{h}^{\max}}}

\begin{document}

\title{{\small{Hadron Collider Physics Symposium (HCP2008),
Galena, Illinois, USA}}\\ 
\vspace{12pt}
Searches for Higgs Bosons beyond the Standard Model at the Tevatron Collider} 

%

\author{F. Filthaut \emph{for the CDF and D0 Collaborations}}
\affiliation{Radboud University and Nikhef, 6525 AJ Nijmegen, The Netherlands}

\begin{abstract}
  The rapidly increasing integrated luminosity collected by the CDF and D0
  detectors at the Tevatron Collider has resulted in a wealth of new results
  from searches for Higgs bosons in the extensions of the Standard Model.
  Tighter limits are set on the parameters governing the Higgs sector in these
  models.
\end{abstract}

\maketitle

\thispagestyle{fancy}


\section{\label{intro}EXTENSIONS OF THE  STANDARD MODEL HIGGS SECTOR}

It is well known that the Standard Model (SM) of particle physics, successful as
it is in describing the strong, electromagnetic, and weak interactions,
cannot be complete. It does not allow for the incorporation of gravitation; and
the mass of its scalar Higgs boson receives quadratic radiative corrections,
making it unnatural for it to be as low as the order of the electroweak symmetry
breaking scale, several hundred \GeV.

The most popular extension of the SM designed to address these issues is
Supersymmetry or SUSY. For each SM particle, it introduces a partner with spin
differing by half a unit but with otherwise the same properties.
This allows for a cancellation of radiative corrections due to
loops of SM particles and their superpartners. But Supersymmetry
must be broken, for otherwise the superpartners' masses would equal those of the
SM particles, and SUSY would have long been discovered. The precise breaking
mechanism depends on the interactions at energy scales inaccessible by the
present-day colliders, and is not known.


A practical approach towards a description of SUSY phenomena at the energy
scales accessible at experiments being conducted or in preparation is to
consider the general \emph{effective} theory obtained upon SUSY
breaking: this is the Minimal Supersymmetric Standard Model (MSSM). Besides
SUSY conserving interactions, it features \emph{soft SUSY breaking} interactions
that still maintain the cancellation of quadratic corrections to scalar masses,
one of the reasons to introduce SUSY in the first place. The MSSM constructed in
this fashion does have one severe drawback, however: it features 105 new and
\emph{a priori} arbitrary parameters. Even if many of these would give rise to
unobserved features such as lepton flavour violation, and are hence constrained
by measurements, the exploration of the full MSSM parameter space is a daunting
task.

In contrast, the same Higgs sector that is at the root of the problems with the
SM is described (at tree level) by a mere two parameters. SUSY requires the
existence of two Higgs doublets (of hypercharge $Y=1$ and $Y=-1$). Upon
electroweak symmetry breaking, this results in five physical Higgs bosons: two
charged ones, $H^{\pm}$, two CP-even neutral ones, $h$ and $H$ (by convention
$m_{h} < m_{H}$), and one CP-odd one, $A$.
At tree level, all masses can be expressed as a function of only two parameters,
usually taken to be $m_{A}$ and the ratio of the two Higgs doublets' vacuum
expectation values $\tan\beta\equiv v_{1}/v_{2}$.

Also the Higgs boson couplings to SM particles can be expressed as a function of
$m_{A}$ and $\tan\beta$. Table~\ref{tab:couplings} shows the neutral Higgs boson
couplings (here, $\alpha$ is not a free parameter but itself depends on
$\tan\beta$ and $m_{A}$); the $H^{\pm}u_{i}d_{j}$ coupling is proportional to
$V_{u_{i}d_{j}}
\left(m_{u_{i}}\cot\beta(1-\gamma_{5})+m_{d_{j}}\tan\beta(1+\gamma_{5})\right)$.
\begin{table}[thb]
  \centering
  \caption{\label{tab:couplings}MSSM Neutral Higgs boson couplings to SM
    particles, relative to the respective SM Higgs boson couplings.}
  \begin{tabular}{|l|c|c|c|}
    \hline
    SM particle type & $h$ coupling & $H$ coupling & $A$ coupling \\
    \hline
    \hline
    up-type quarks & $\frac{\textstyle\cos\alpha}{\textstyle\sin\beta}$ &
    $\frac{\textstyle\sin\alpha}{\textstyle\sin\beta}$ & $\cot\beta$ \\
    \hline
    down-type quarks, $\ell^{\pm}$ &
    $-\frac{\textstyle\sin\alpha}{\textstyle\cos\beta}$ &
    $\frac{\textstyle\cos\alpha}{\textstyle\cos\beta}$ & $\tan\beta$ \\
    \hline
    W and Z bosons & $\sin(\beta-\alpha)$ & $\cos(\beta-\alpha)$ & 0 \\
    \hline
  \end{tabular}
\end{table}

A remarkable feature is that (again at tree level) $m_{h} < m_{Z}$. This sole
fact would be enough to exclude the MSSM, were it not for the fact that
radiative corrections relax these constraints. In principle, these radiative
corrections involve the whole MSSM, but the number of parameters contributing
substantially is still small. As a general matter, a constraint $m_{h}\lesssim
135 \GeV$ remains.

The radiative corrections affect the Higgs bosons' couplings as well as their
masses. Rather than considering the whole MSSM parameter space, several
representative \emph{scenarios} are typically considered that evade the bounds
set by searches for Higgs bosons conducted at the LEP $e^{+}e^{-}$ Collider.
Specifically, the scenarios considered here are \mhmax\ scenario (with MSSM
parameters tuned to obtain the highest possible $m_{h}$) and the no-mixing
scenario (with parameters tuned to switch off the mixing between the
$\tilde{t}_{1}$ and $\tilde{t}_{2}$ superpartners of the top
quark)~\cite{ref:scenarios}. Within these scenarios, the Higgsino mass parameter
$\mu$ is in principle fixed; however, given the sensitivity to this parameter,
both $\mu=+200 \GeV$ and $\mu=-200 \GeV$ are considered in some of the analyses
below.

\section{\label{neutral}NEUTRAL HIGGS BOSONS}

\subsection{\label{neutral-general}Generalities}

Previous searches at LEP for neutral Higgs bosons have excluded a large region
of the ($m_{A}$,$\tan\beta$) parameter space where the lightest Higgs boson
becomes sufficiently light to be accessible at LEP. In particular, the low
$\tan\beta$ region has been excluded up to large values of $m_{A}$.

The searches conducted at the Tevatron Collider are largely complementary to
those earlier searches. At large values of $\tan\beta$, the enhanced couplings
to $b$ quarks and $\tau$ leptons lead to significantly increased production
cross sections. This fact, in combination with the particular signatures of
$b$-quark jets and $\tau$ leptons, is exploited below.

As a consequence of large $\tan\beta$, the masses of the $A$ and $H$ bosons (or
$A$ and $h$, for low values of $m_{A}$) become almost degenerate (their mass
difference becomes less than the corresponding detector resolution). This allows
for a significant simplification in the analyses: rather than attempting to
identify the two nearby resonances, they are simply lumped together, and denoted
as $\phi$ below. By comparison, the contribution from the ``third'' neutral
Higgs boson becomes negligible.

\subsection{\label{bb}\boldmath$\phi\rightarrow b\bar{b}$}

The $b$ quark being the heaviest SM isospin $-\frac{1}{2}$ fermion, the
$\phi\rightarrow b\bar{b}$ branching fraction is the largest one for high values
of $\tan\beta$, becoming as high as $\sim$ 0.9. As a consequence, it constitutes one of the
dominant search channels. Unfortunately, this decay cannot be identified
inclusively, due to an overwhelming QCD $b\bar{b}$ production
background. Instead, one looks for the final state in which the $\phi$ is
radiated off an initial state $b$ quark\footnote{Here and in the following,
  charge conjugated final states are implied as well.}.
A significant ($\sim \tan^{2}\beta$) cross section enhancement occurs also in
this final state; the identification of a third $b$ jet, typically of lower
$p_{T}$, suppresses greatly the QCD background.

The CDF Collaboration has searched for the $b\phi(\rightarrow b\bar{b})$ channel
in a data sample corresponding to an integrated luminosity of 1.9 fb$^{-1}$,
requiring all jets in a three-jet sample to be $b$-tagged by a displaced vertex
algorithm~\cite{ref:cdf-hbb}.

Even this triple-tagged sample is dominated by QCD backgrounds. Each jet may be
a genuine $b$-quark jet, or a fake (charm or light-flavour) jet. The two leading
jets, hypothesized to result from the $\phi$ decay, are treated on an equal
footing. Therefore, considering only those final states containing at least two
genuine $b$-quark jets leads to five background categories: $bbb$, $bbc$, $bbq$,
$bcb$, and $bqb$ (here, $q$ denotes a light-flavour jet).

The distribution of the invariant mass $m_{12}$ of the two leading jets for the
signal is taken from simulation. For the backgrounds no adequate simulation
exists, and templates for each of the above categories are obtained starting
from a data sample containing at least two $b$ tags. For the first three
categories, this starts from a $bbj$ sample to which parametrized efficiencies
for tagging the third jet ($b$, $c$, or $q$) are applied; for the $bqb$
category, a similar procedure is applied to events with the third and one of the
two leading jets already tagged.
An additional correction, derived from simulation, is applied to account for the
different processes contributing to final states with more than two $b$
jets. Finally, the $bcb$ distribution is assumed to look like the $bbb$ one, but
with only a subset of processes contributing; another simulation-derived
correction is applied to account for this.

\begin{figure}[htb]
  \centering
  \includegraphics[width=0.47\textwidth]{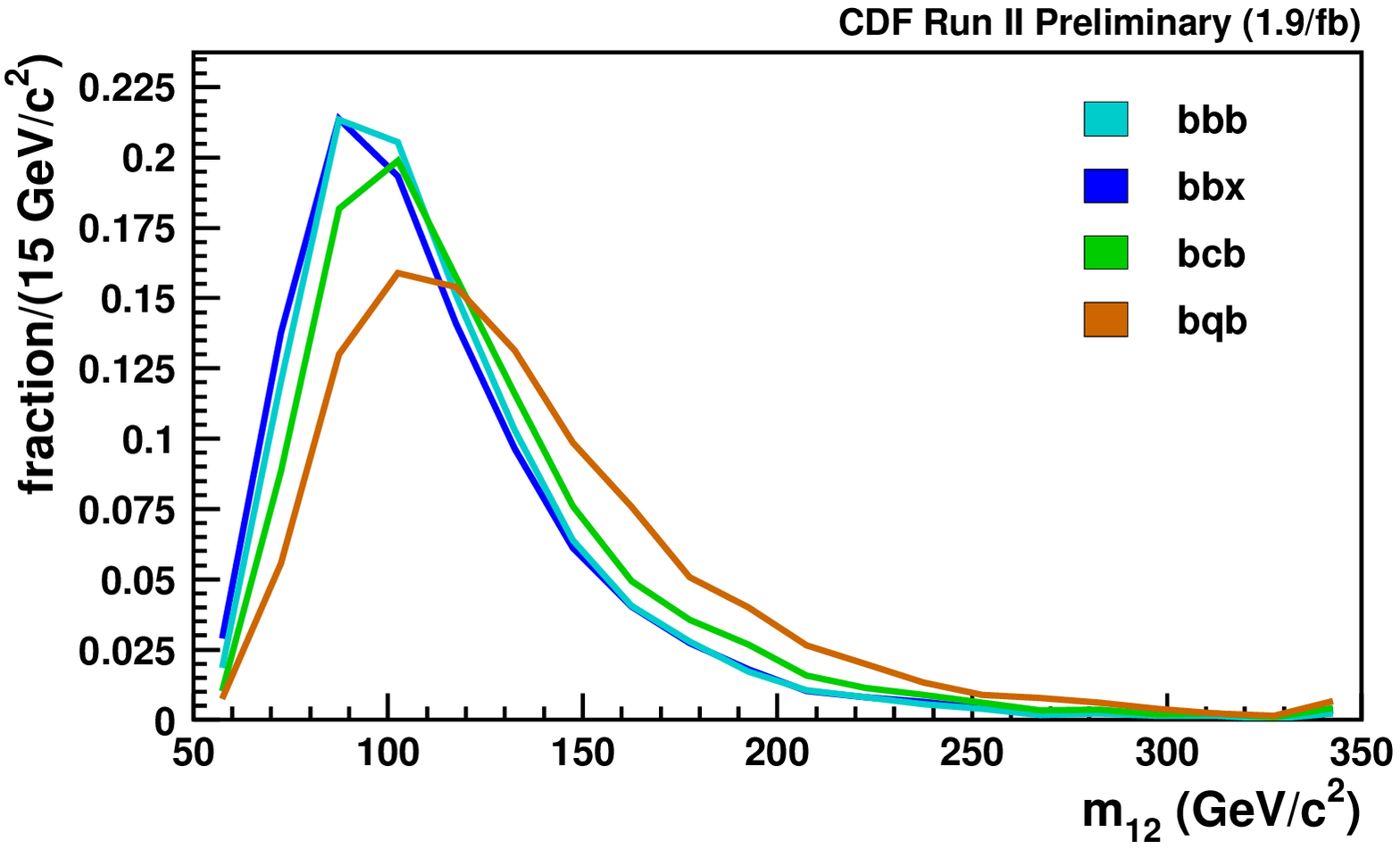}
  \includegraphics[width=0.47\textwidth]{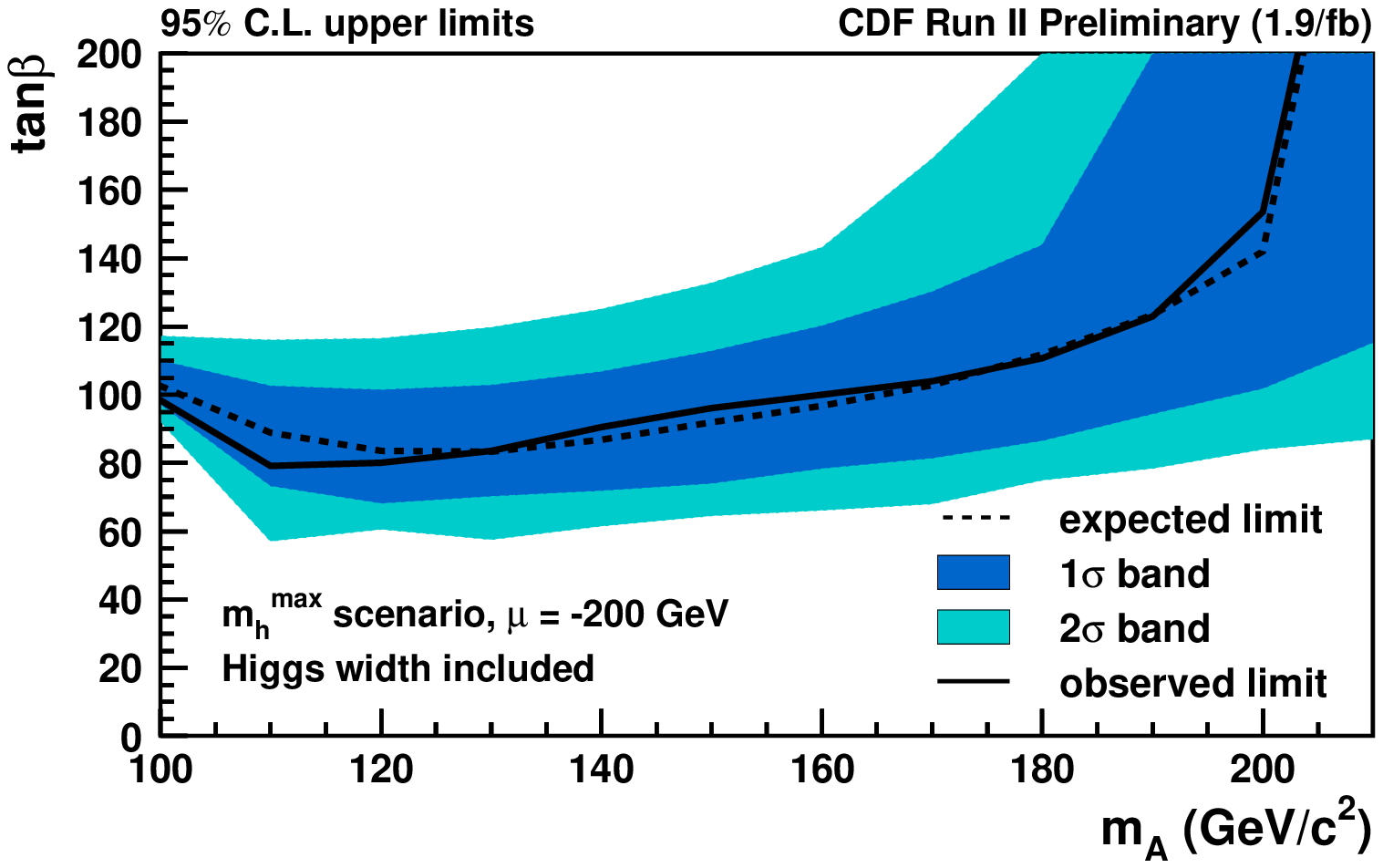}
  \caption{\label{cdf-hbb}Left: invariant mass distributions for various
    background sources in the CDF $\phi\rightarrow b\bar{b}$ analysis. Right:
    limits in the ($m_{A}$, $\tan\beta$) plane in the \mhmax\ scenario.}
\end{figure}

The normalization of the distributions obtained in this fashion are adjusted to
fit the data distribution. However, as can be observed from Fig.~\ref{cdf-hbb},
some of the background templates cannot be distinguished. Therefore, a second
variable involving the vertex masses (the invariant mass of all tracks, assuming
they are $\pi^{\pm}$, attached to the displaced vertices) is constructed. This
is done starting from the same data samples above, and uses the same information
in addition to known vertex mass distributions for different flavours.

For each hypothesized value of $m_{\phi}$, two template fits are then carried out
jointly to the two distributions, one with and one without the assumed
signal. No significant improvement in the fit likelihood is observed upon
inclusion of a signal component; limits on a possible signal component are set
in a modified frequentist framework based on~\cite{ref:frequentist}.

A complication is that for sufficiently high values of $\tan\beta$, the neutral
Higgs boson decay width exceeds the experimental invariant mass resolution. The
resulting broadening of the signal $m_{12}$ distribution reduces the
discrimination power between signal and backgrounds, and needs to be taken into
account explicitly.

A similar analysis has been carried out by the D0 Collaboration on a 1 fb$^{-1}$
dataset~\cite{ref:d0-hbb}. An artificial neural network $b$-tagging algorithm
exploiting lifetime information was employed. Rather than using a second
discriminating variable to determine the various background components, multiple
$b$-tagging criteria (with known efficiencies) are used to determine the
composition of the double-tagged sample.

\begin{figure}[htb]
  \centering
  \includegraphics[width=0.44\textwidth]{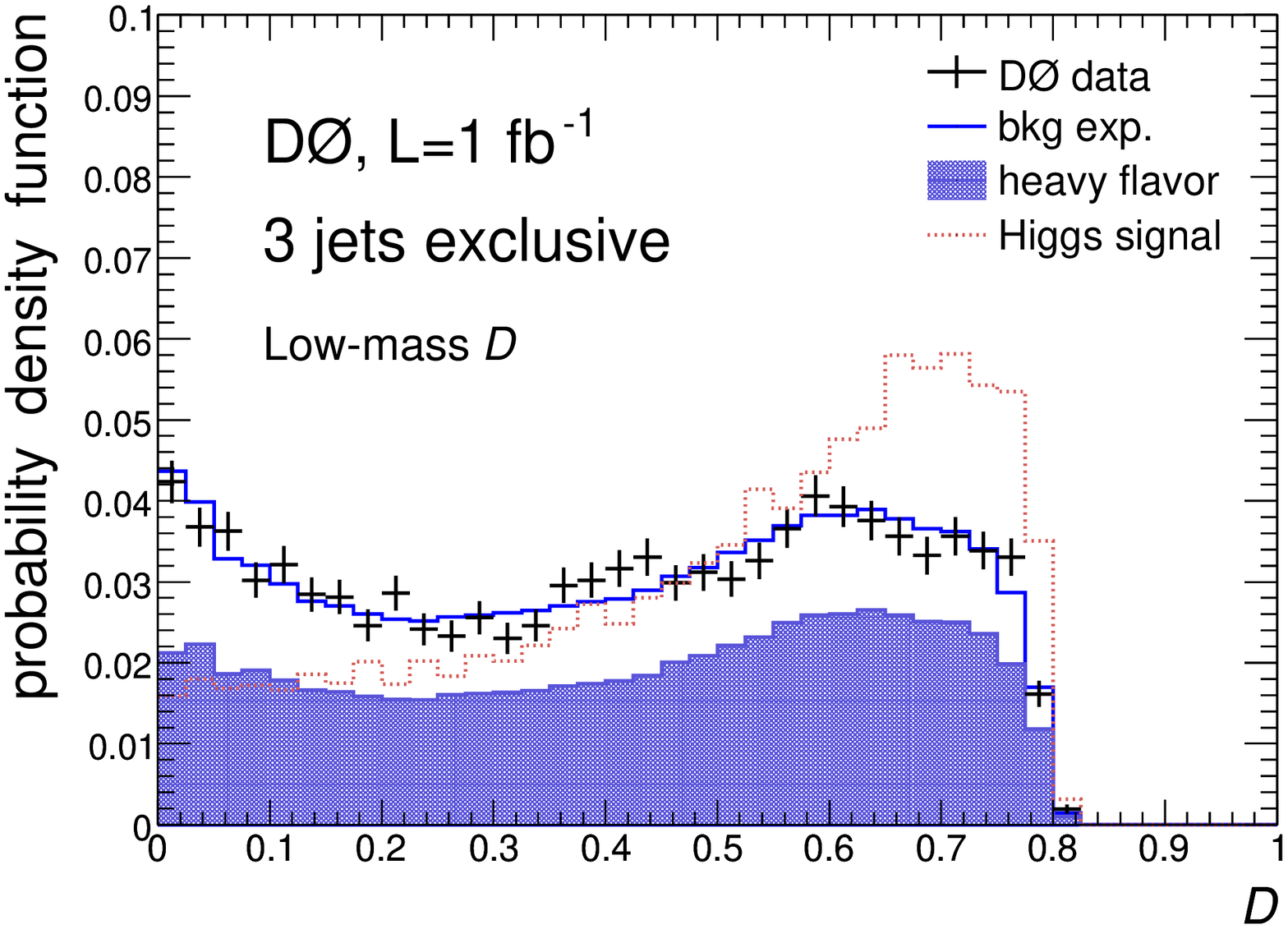}
  \includegraphics[width=0.17\textwidth]{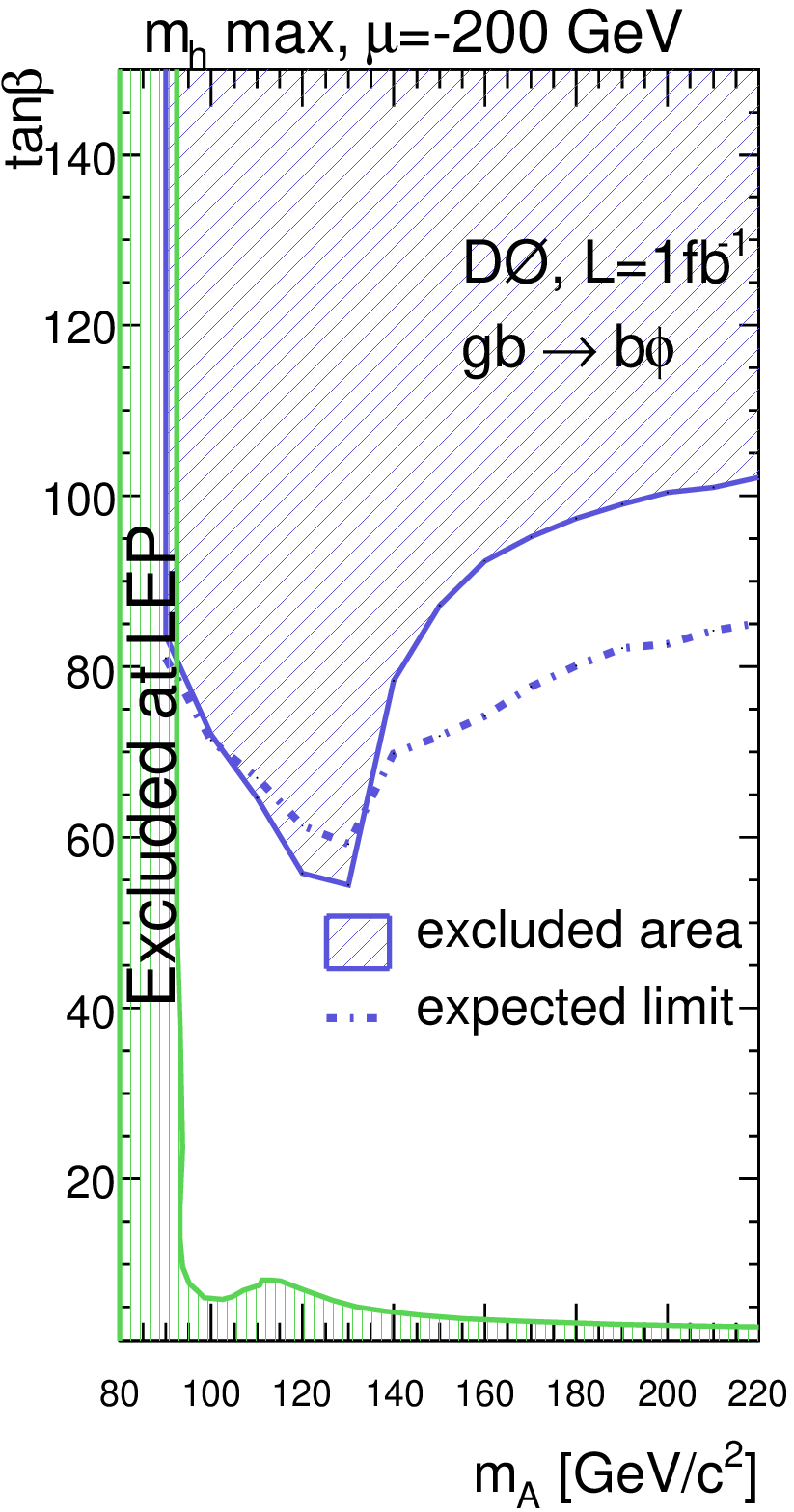}
  \includegraphics[width=0.17\textwidth]{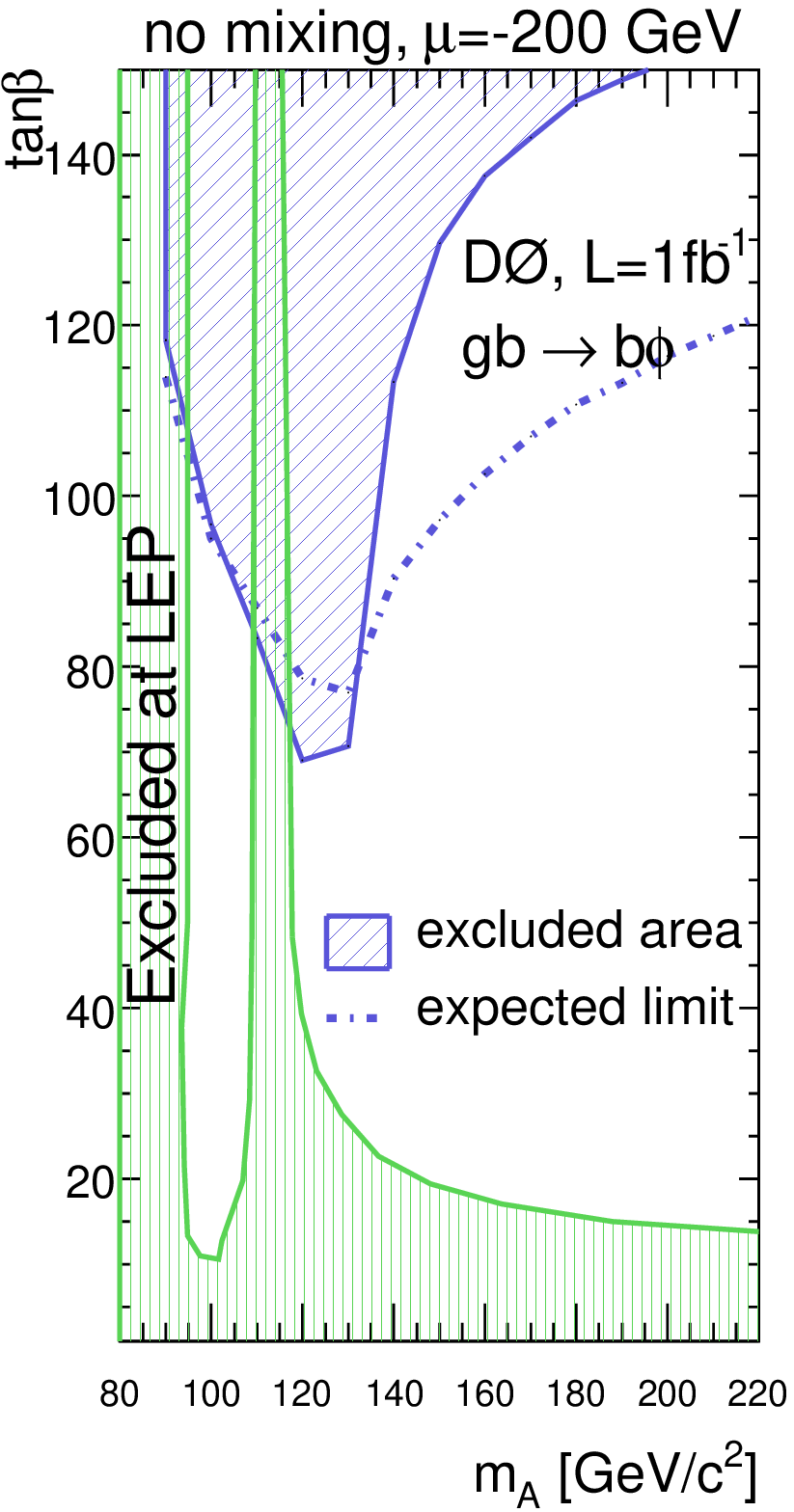}
  \includegraphics[width=0.17\textwidth]{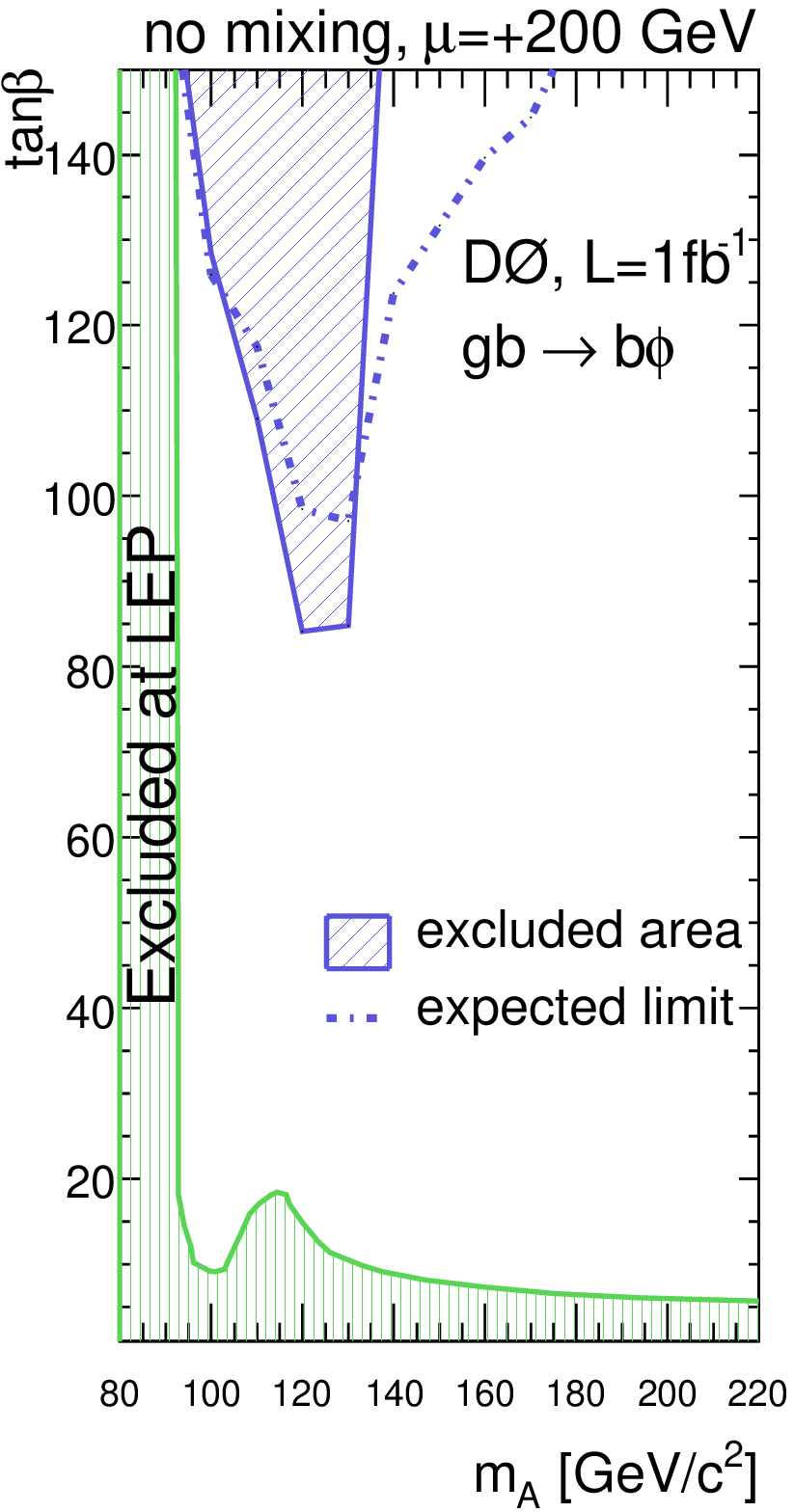}
  \caption{\label{d0-hbb}Left: low-mass kinematic discriminant distribution in
    the D0 $\phi\rightarrow bb$ analysis. Right: limits in the ($m_{A}$,
    $\tan\beta$) plane in different scenarios.}
\end{figure}

A second discriminating variable, based on kinematic information, is
subsequently used nevertheless, to suppress further the QCD backgrounds. Two
discriminants $D$ are constructed, one combining low-mass (90--130 \GeV) and one
combining high-mass (130--220 \GeV) simulated Higgs signal samples. The
background predictions are obtained, as a function of both the invariant mass
and the discriminant, by applying a correction derived from simulation to the
double-tagged sample. Fig.~\ref{d0-hbb} compares the low-mass discriminant
distribution integrated over all invariant masses, with its prediction; a good
agreement is observed for the background dominated low-$D$ region. The final
background predictions are obtained by selecting the high-$D$ region. The whole
analysis is applied jointly to exclusive three-, four-, and five-jet data
samples. Again no significant excess of data over predictions is observed, and a
modified frequentist method based on~\cite{ref:frequentist} is used to set
limits on a possible signal. The results are also shown for both the no-mixing
and \mhmax\ scenarios in Fig.~\ref{d0-hbb}. A slight excess, not enough to be
qualified as a signal, is observed at masses between 150 and 200 \GeV, leading
to a reduced exclusion region compared to the predictions.

\subsection{\label{bb}\boldmath $\phi\rightarrow \tau^{+}\tau^{-}$}

For high $\tan\beta$, the branching fraction for the decay $\phi\rightarrow
\tau^{+}\tau^{-}$ is approximately 0.1. The drawback of this lower branching
fraction is offset, however, by the absence of irreducible QCD background. The
relative cleanliness of this decay channel allows for an inclusive analysis,
where the $\phi$ can be produced singly, through gluon fusion
$gg\rightarrow\phi$.

A new D0 analysis has been performed on $e\tau_{h}$, $\mu\tau_{h}$, and $e\mu$
final states (here, only the $\tau$ decay products are indicated and neutrinos
omitted; $\tau_{h}$ indicates the $\tau$ hadronic decay mode) in a 1 fb$^{-1}$
dataset~\cite{ref:d0-htautau}. Hadronic $\tau$ decays are classified according
to their charged track and EM cluster multiplicities, and are selected
efficiently using artificial neural networks.

The signal is characterized by two high-$p_{T}$ $\tau$ leptons. To reject
$W(\rightarrow\ell\nu)+$jets events with a jet misidentified as a $\tau_{h}$,
the transverse mass reconstructed from lepton and the missing transverse
momentum vector (interpreted as a single neutrino) must be below 40 \GeV\ in the
$\mu\tau_{h}$ channel and 50 \GeV\ in the $e\tau_{h}$ channel. Backgrounds from
$t\bar{t}$ events are suppressed by vetoeing events with high jet activity.

The irreducible background from $Z\rightarrow\tau^{+}\tau^{-}$ remains and is
estimated using simulation. Signal events are recognized as an excess over
background predictions in the distribution of the \emph{visible mass}
$m_{\mbox{\small vis}}$, computed as the invariant mass of the visible $\tau$
decay products and the missing momentum four-vector approximated as
$\slashchar{p} \equiv \left(\sqrt{\slashchar{E}_{x}^{2}+\slashchar{E}_{y}^{2}},
  \slashchar{E}_{x}, \slashchar{E}_{y}, 0\right)$.

\begin{figure}[htb]
  \centering
  \includegraphics[width=0.8\textwidth]{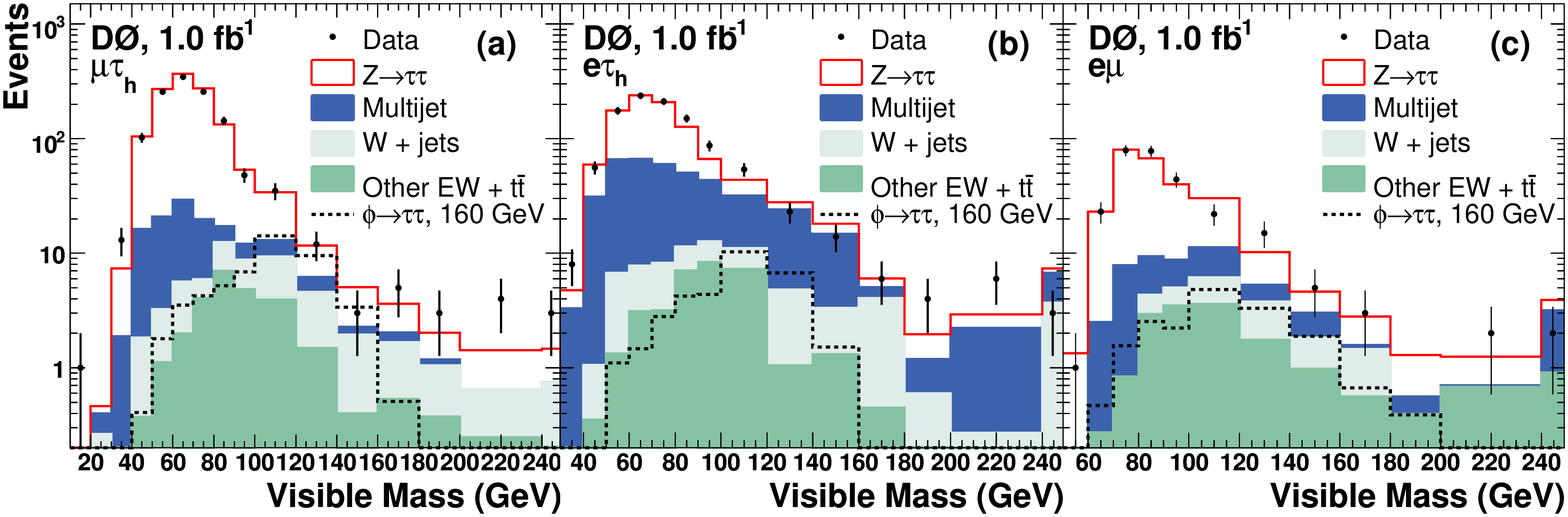}
  \includegraphics[width=0.32\textwidth]{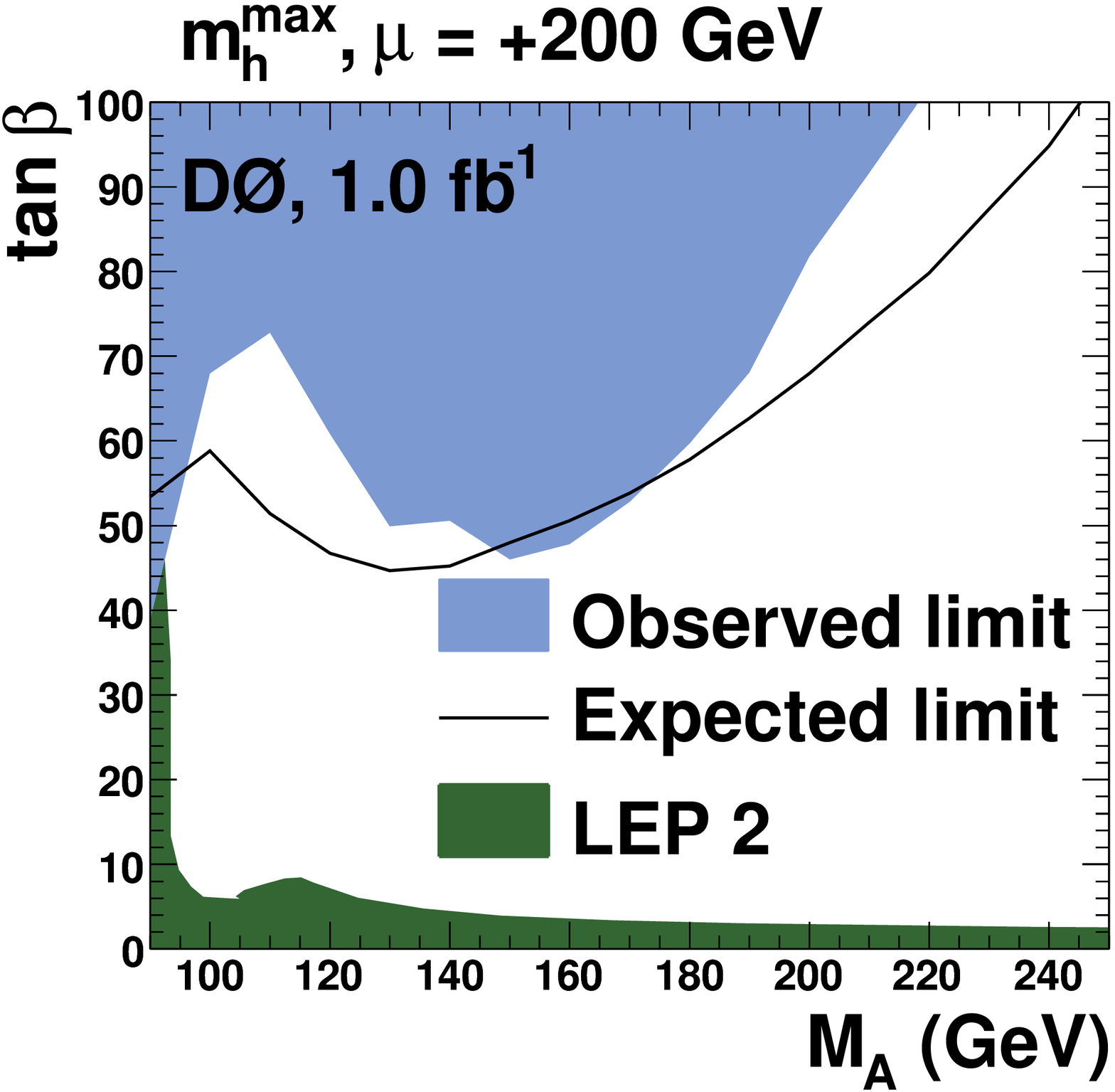}
  \includegraphics[width=0.32\textwidth]{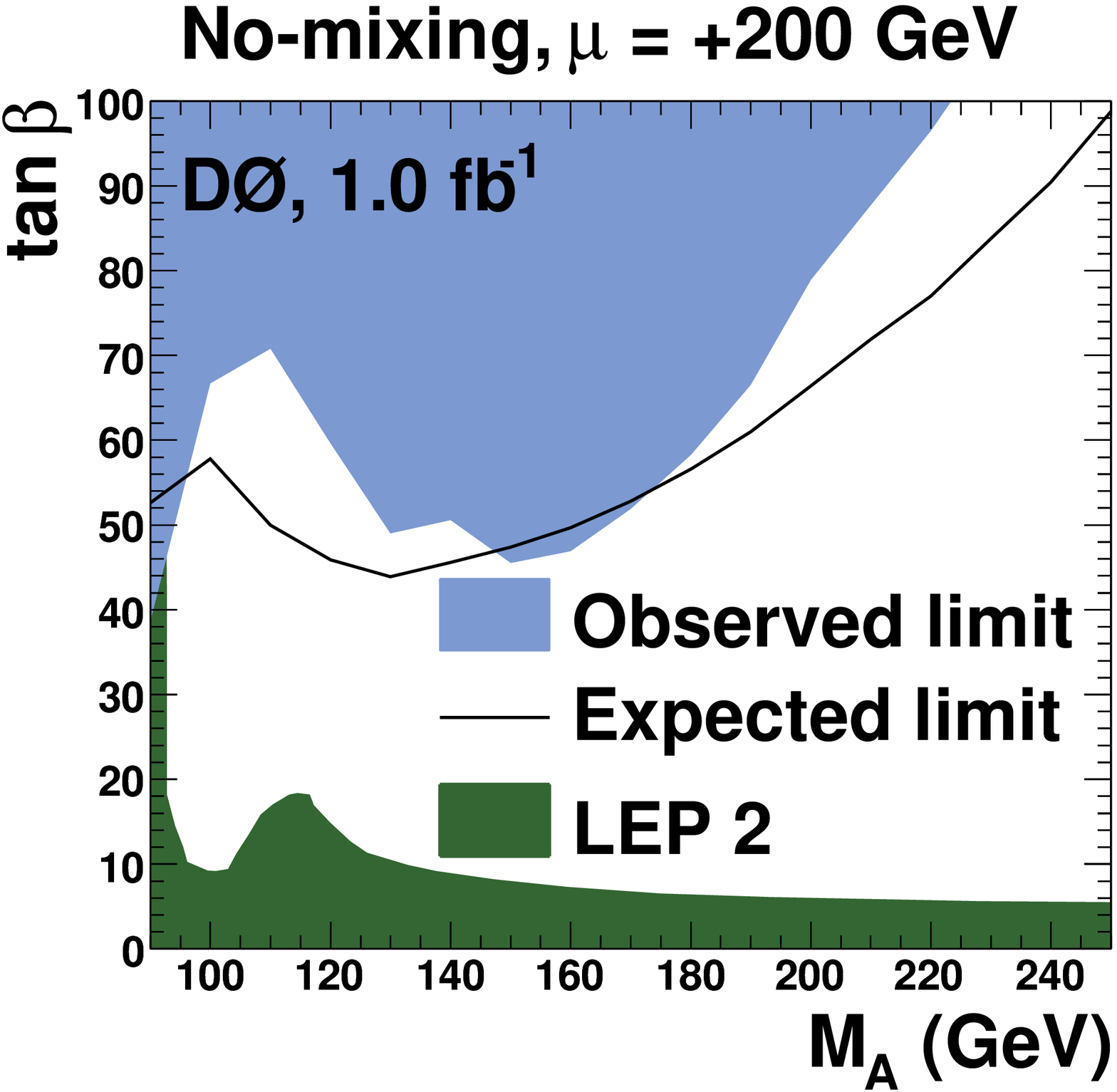}
  \caption{\label{d0-tautau-mvis}Top: visible mass distribution in the three
    channels considered in the D0 $\phi\rightarrow\tau^{+}\tau^{-}$ search.
    Bottom: limits in the ($m_{A}$, $\tan\beta$) plane obtained in the \mhmax\
    (left) and no-mixing scenarios.}
\end{figure}

The visible mass distribution is shown in Fig.~\ref{d0-tautau-mvis} for the
three channels considered. The expected signal due to a 160 \GeV\ Higgs boson is
also indicated. No significant excess is observed, and limits are again set
using the modified frequentist framework. It should be pointed out that here,
due to the worse experimental resolution caused by the escaping neutrinos, the
effects of the finite $\phi$ decay width are very modest.

\begin{figure}[htb]
  \centering
  \includegraphics[height=0.2\textheight]{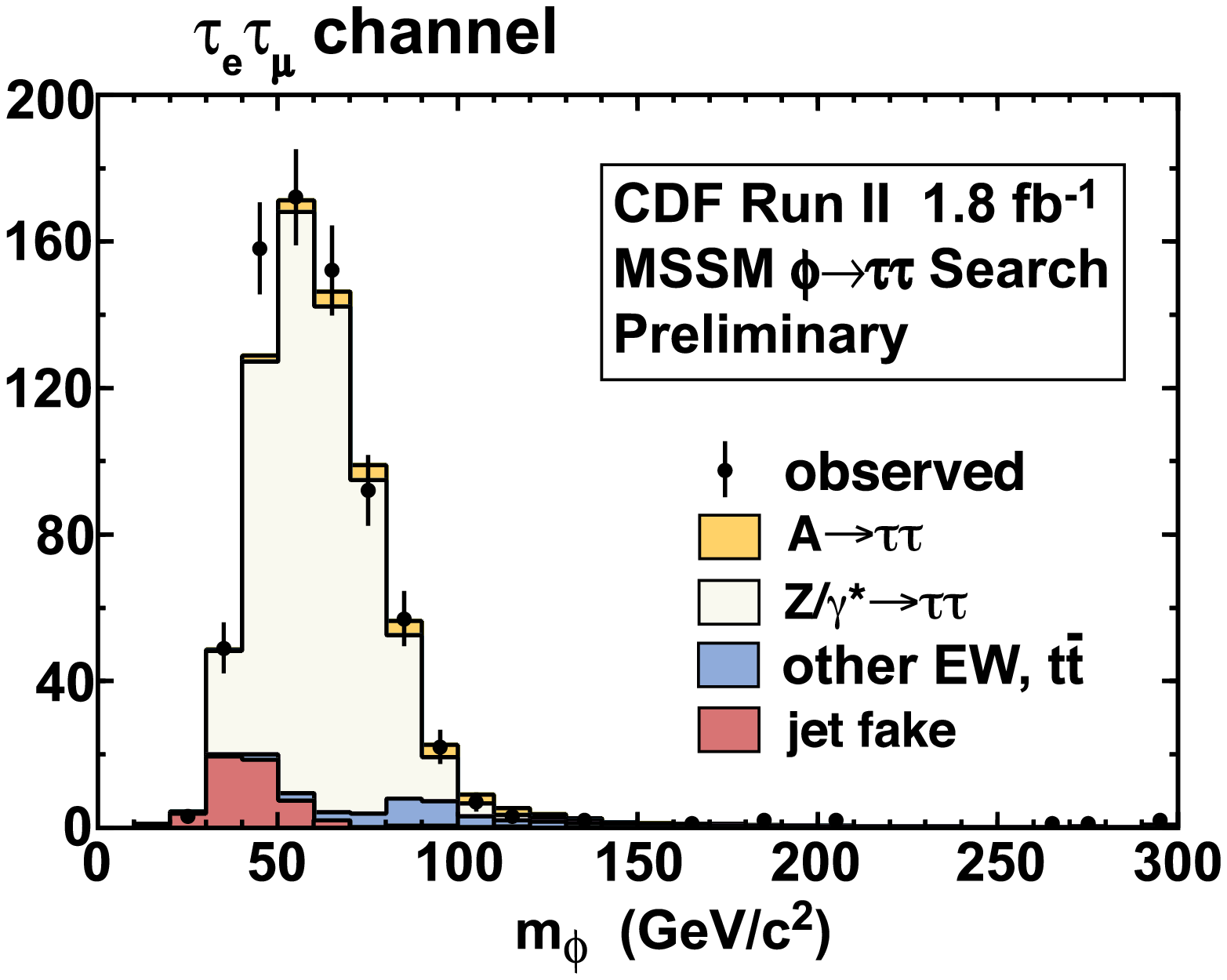}
  \includegraphics[height=0.2\textheight,clip]{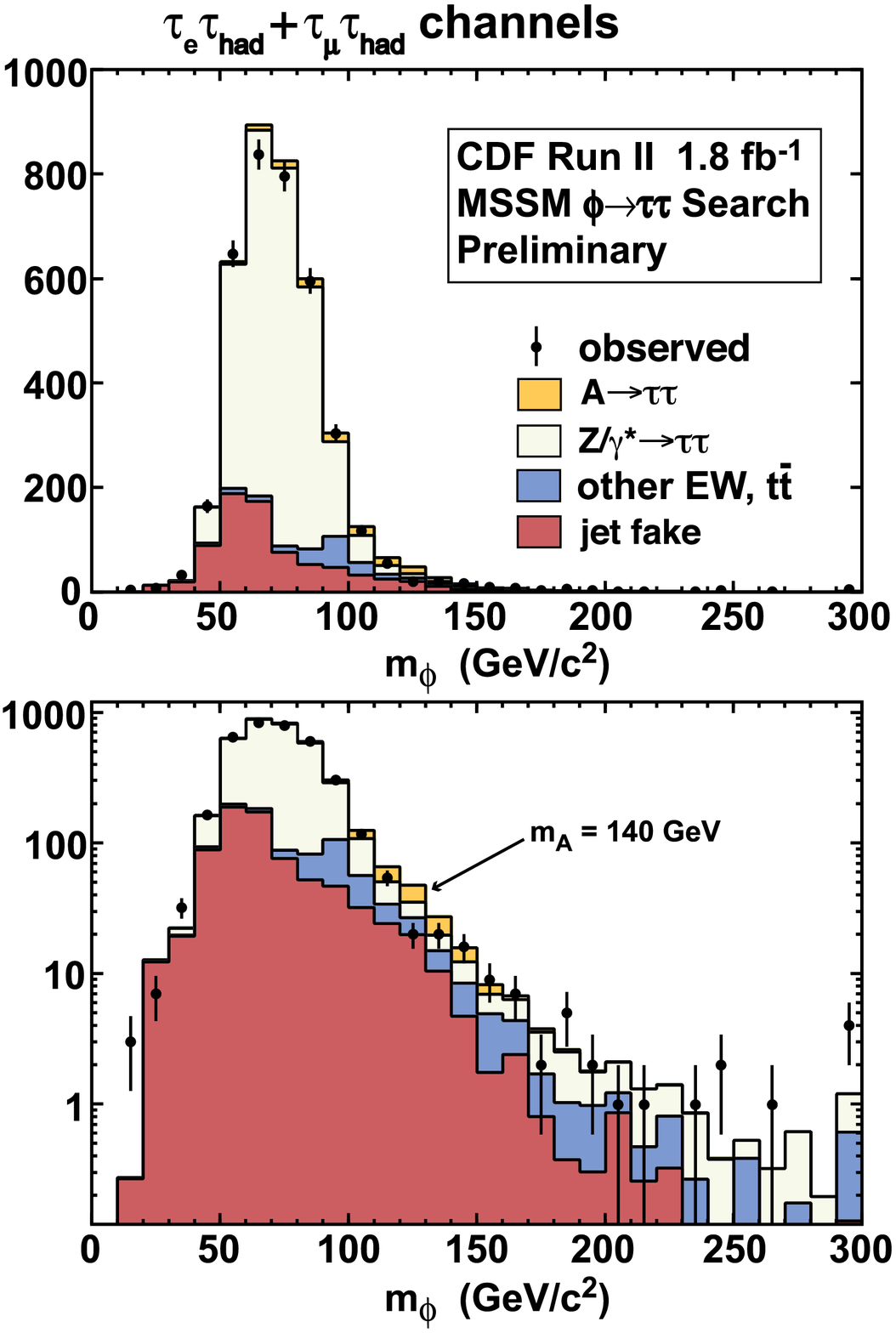}
  \includegraphics[width=0.39\textwidth]{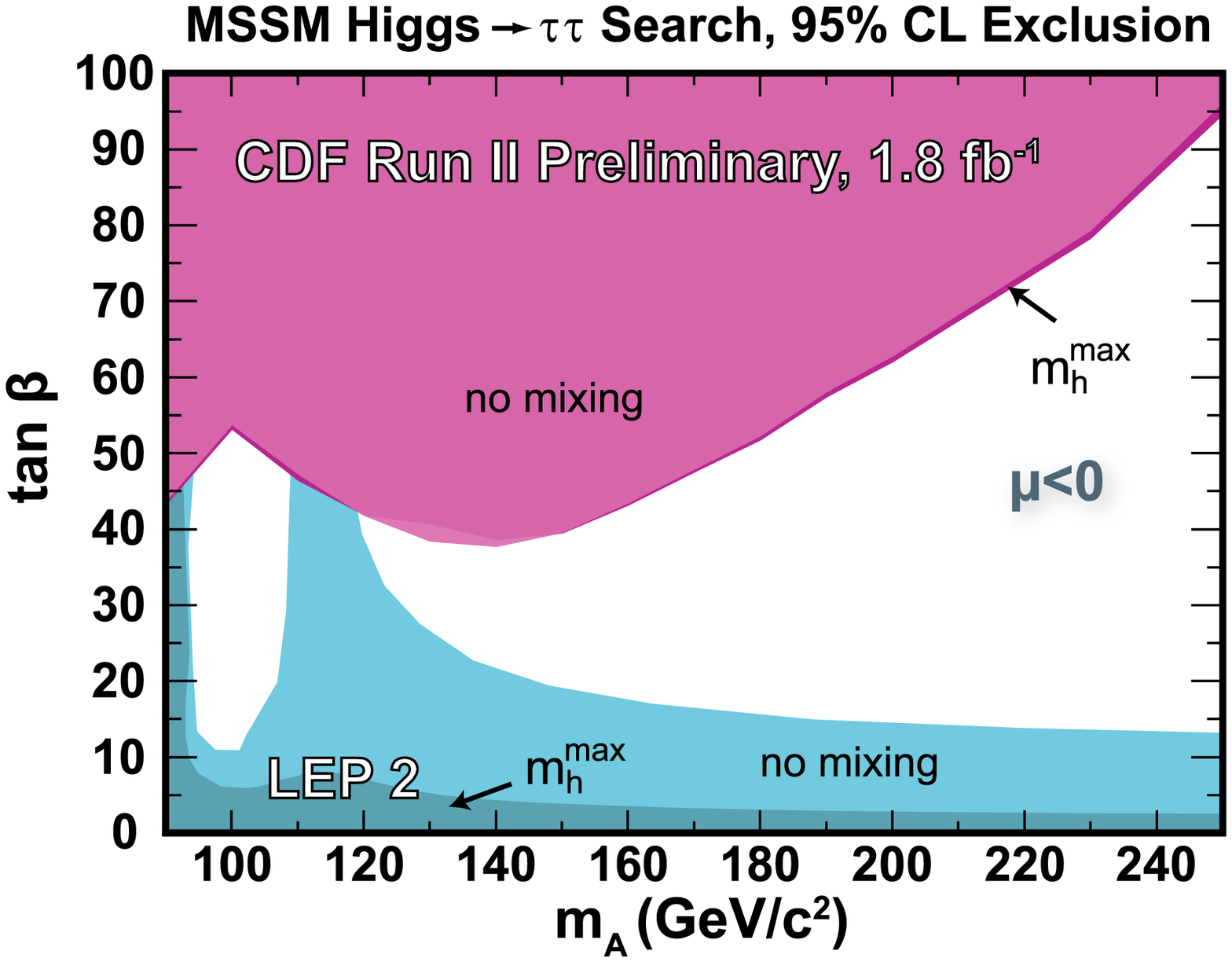}
  \includegraphics[width=0.39\textwidth]{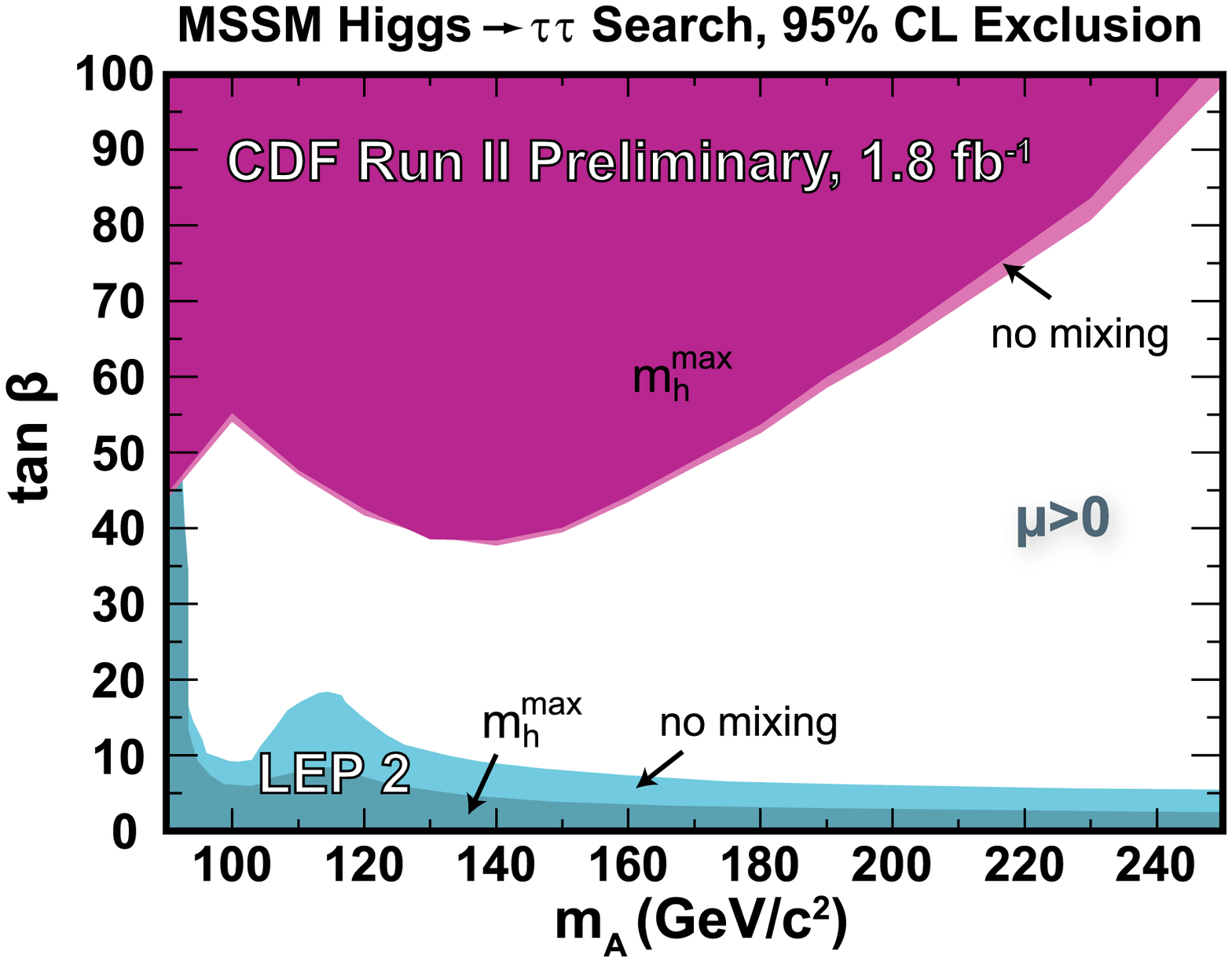}
  \caption{\label{cdf-tautau}Top: visible mass distribution in the $e\mu$
    channel (left) and in the $e\tau_{h}+\mu\tau_{h}$ channels (right) in the
    CDF $\phi\rightarrow\tau^{+}\tau^{-}$ search.
    Bottom: limits in the ($m_{A}$, $\tan\beta$) plane, for $\mu<0$ (left)
    and $\mu>0$ (right).} 
\end{figure}

A very similar CDF analysis has been carried out on a 1.8 fb$^{-1}$
dataset~\cite{ref:cdf-htautau}. An earlier analysis, performed on 1 fb$^{-1}$,
yielded a small excess that could be attributed to a Higgs boson of mass
$m_{\phi} \approx 140 \GeV$. With the increased statistics considered here, a
good overall agreement between data and background predictions is obtained. This
analysis now sets the tightest limits in the ($m_{A}$, $\tan\beta$) plane, as
displayed in Fig.~\ref{cdf-tautau}.

\subsection{\label{fermiophobic}Fermiophobic Higgs Bosons}

Besides the reasonably ``standard'' no-mixing and \mhmax\ scenarios, alternative
scenarios are possible where the Higgs boson couplings to fermions are small; an
example is the ``small $\alpha_{\textnormal{\scriptsize eff}}$''
scenario~\cite{ref:scenarios}, in which radiative corrections conspire to yield
small effective couplings to down-type quarks and charged leptons.

The D0 Collaboration has considered a \emph{fermiophobic} Higgs model, in which
all couplings to fermions are suppressed but the couplings to gauge bosons are
unmodified. In a 2.3 fb$^{-1}$ dataset, a search was performed for the decay
$h\rightarrow\gamma\gamma$~\cite{ref:d0-fermiophobic}. This would be the main
decay mode for relatively light Higgs bosons, $m_{h} \lesssim 100 \GeV$.

Backgrounds to this process originate from QCD direct $\gamma\gamma$ events, as
well as $\gamma+$jet and di-jet events in which one or both jets, respectively,
are mis-identified as photons. The mis-identification contribution is suppressed
using a neural network employing variables based on calorimeter shower shapes;
the network performance has been measured in $Z\rightarrow e^{+}e^{-}$ events
and samples of photon-like QCD jets. Starting from a loose preselection, the
identification of events where neither, one, or two of the photon candidates
satisfy a tight neural network cut allows to determine the contributions from
the individual backgrounds. The irreducible $\gamma\gamma$ background estimated
in this way is consistent with the result obtained from a NLO calculation. The
contribution from $Z\rightarrow e^{+}e^{-}$ events where both electrons are
identified as photons, is estimated from simulation.

\begin{figure}[htb]
  \centering
  \includegraphics[width=0.47\textwidth,clip=]{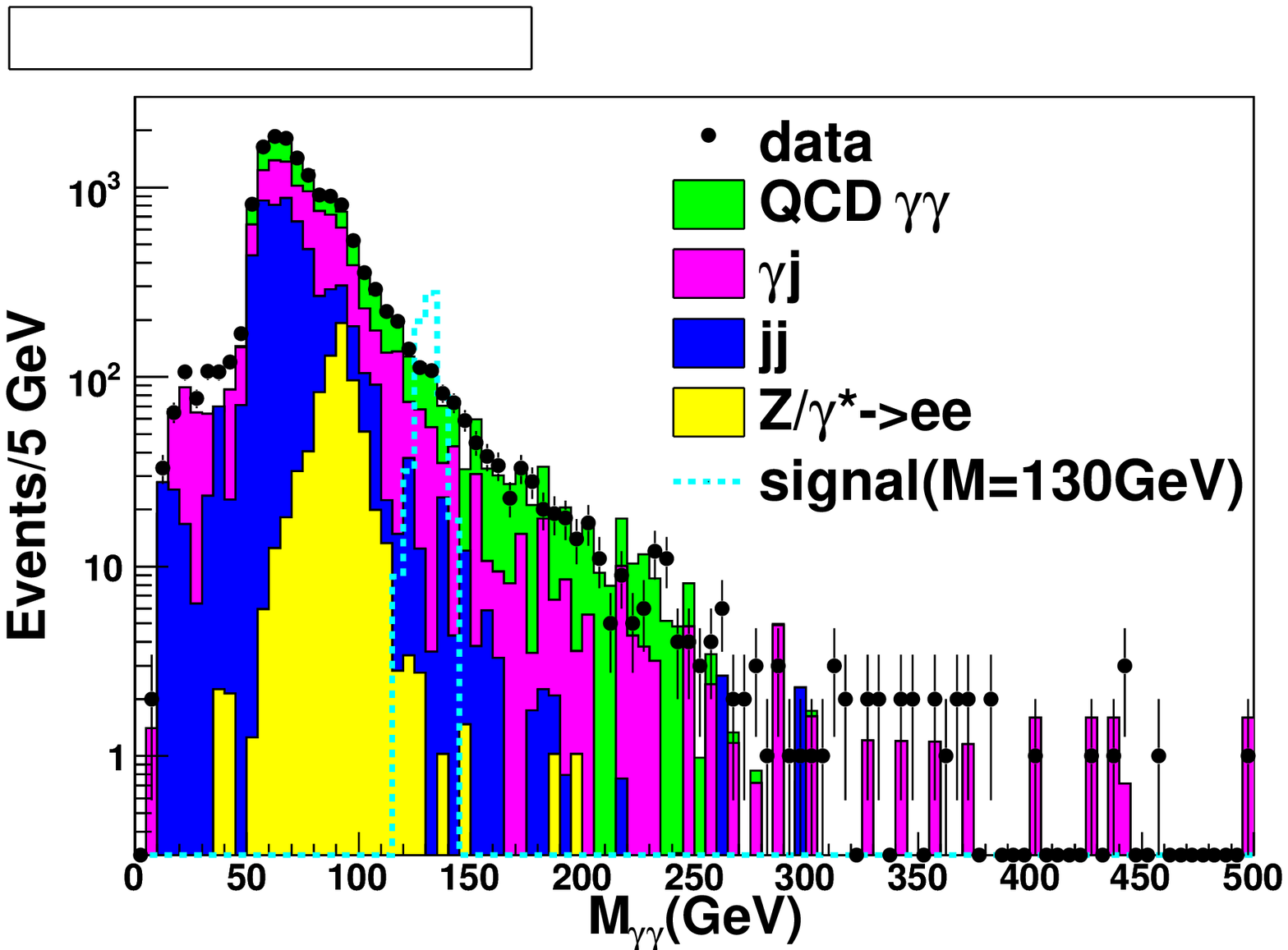}
  \includegraphics[width=0.43\textwidth]{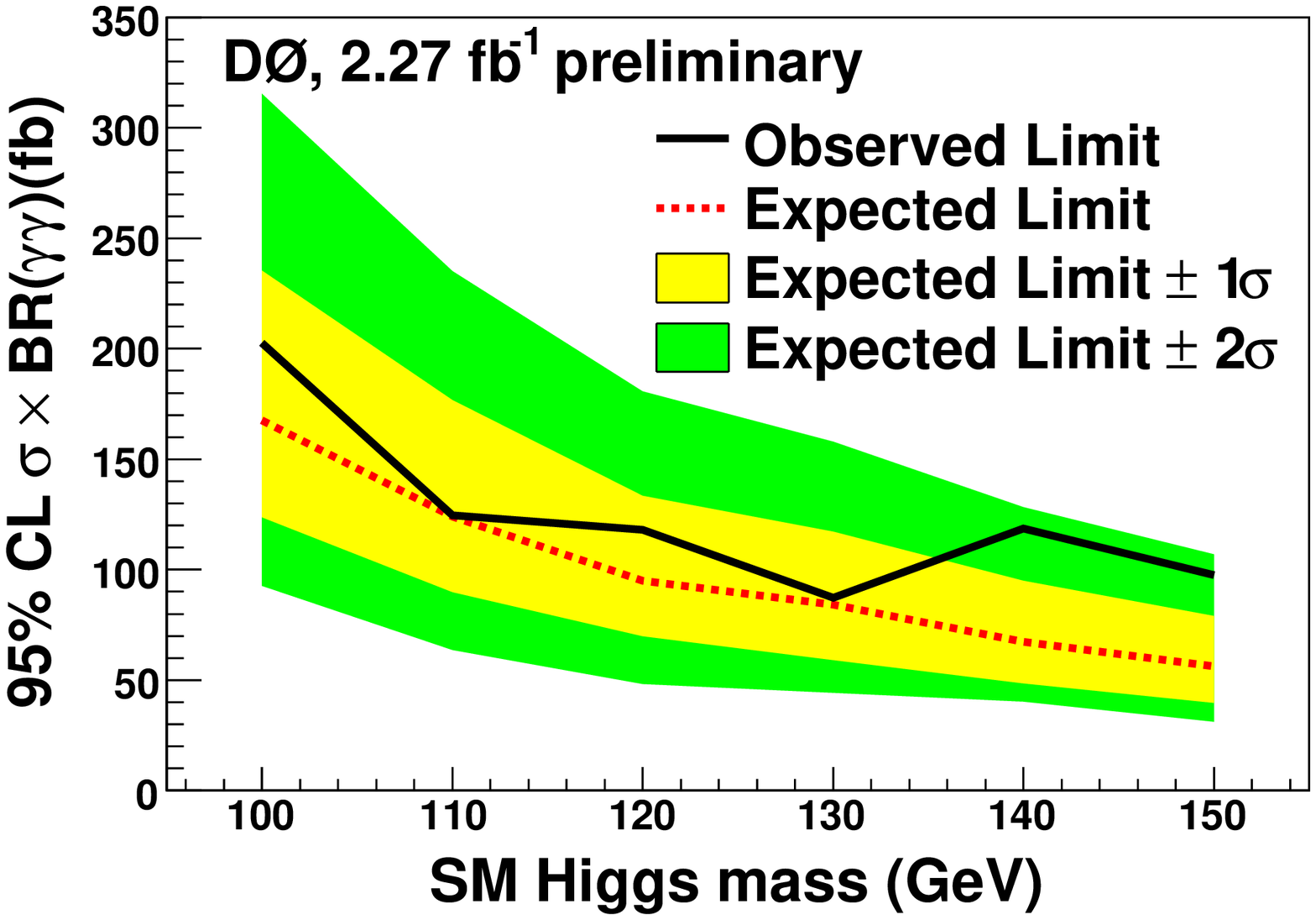}
  \caption{\label{d0-gg}Left: $\gamma\gamma$ invariant mass distribution in the
    D0 $h\rightarrow\gamma\gamma$ search. Right: cross section limit as a
    function of $m_{h}$.}
\end{figure}

The $\gamma\gamma$ invariant mass distribution is shown in Fig.~\ref{d0-gg},
along with the signal of a 130 \GeV\ Higgs boson. No excess over predictions is
observed, and limits are set on the possible cross section for the production of
such events. The limits are also shown in Fig.~\ref{d0-gg}.

For higher values of $m_{h}$, the $h\rightarrow W^{+}W^{-}$ decay mode becomes
dominant. The CDF Collaboration has exploited this by using a SM Higgs boson
search in the channel $W^{\pm}h\rightarrow W^{\pm}W^{+}W^{-}$ in a 1.9 fb$^{-1}$
dataset, and interpreting its result in the context of a search for fermiophobic
Higgs bosons~\cite{ref:cdf-fermiophobic}. The main difference with the
corresponding SM search is that the branching fraction for this Higgs boson
decay mode is relatively high even for lower values of $m_{h}$.

\begin{figure}[htb]
  \centering
  \includegraphics[width=0.4\textwidth]{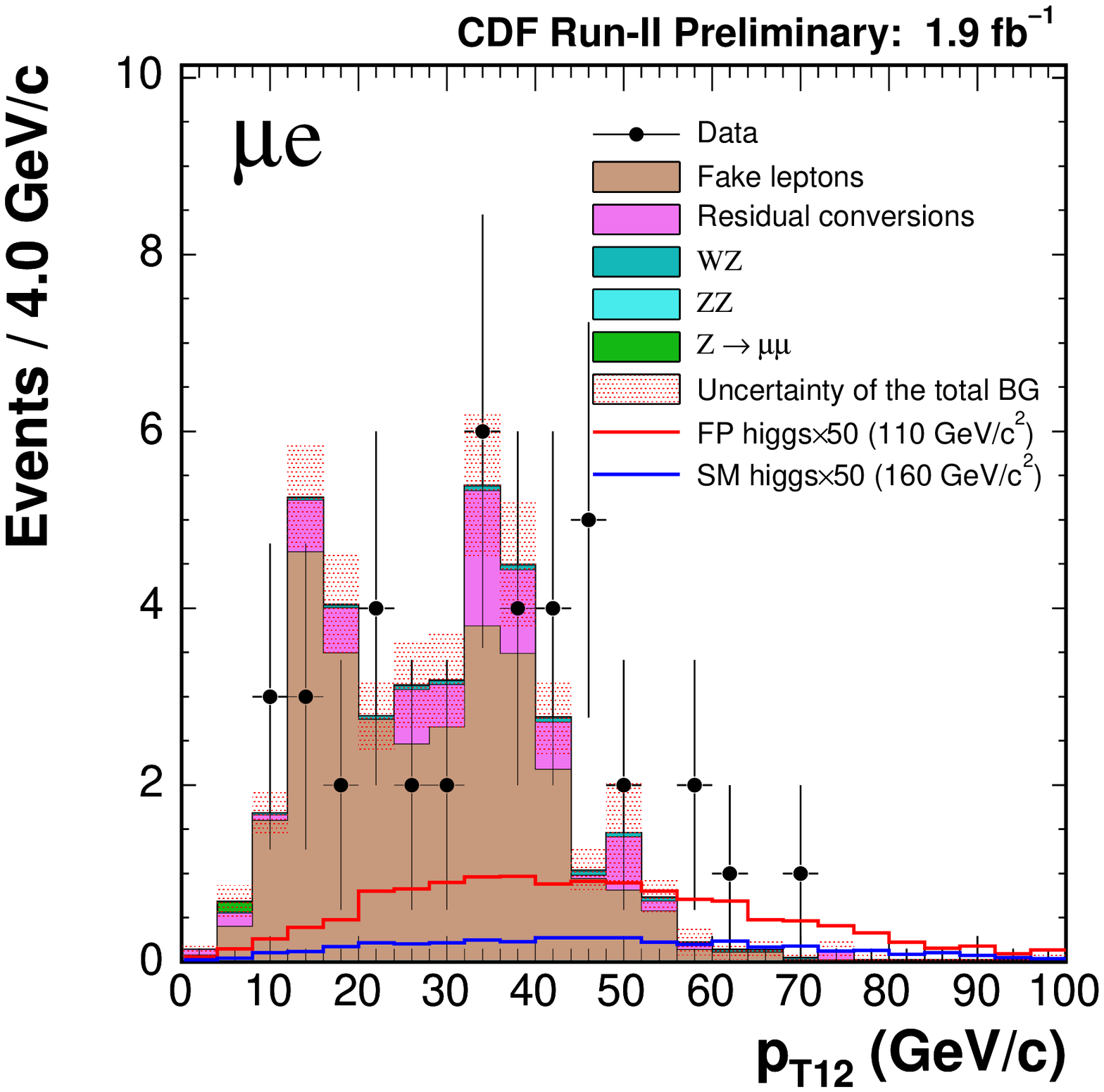}
  \includegraphics[width=0.4\textwidth]{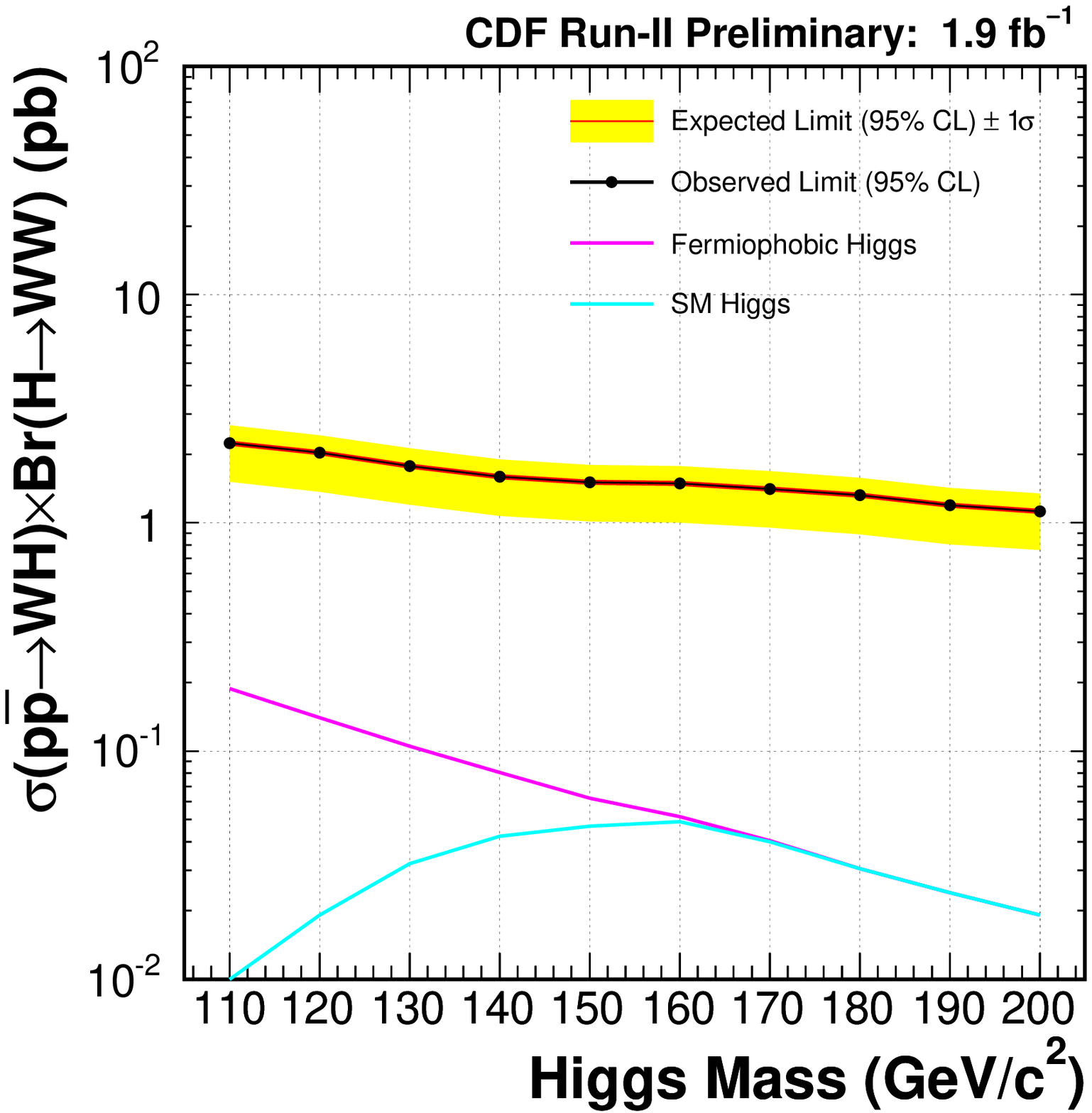}
  \caption{\label{cdf-www}Left: transverse momentum of the dilepton system for the
    $\mu^{\pm}e^{\mp}$ channel in the CDF $Wh\rightarrow WWW$ search. Right: 
    resulting cross section limits as a function of $m_{h}$, compared with the
    predictions for SM and fermiophobic Higgs boson production.}
\end{figure}

The requirement of two isolated leptons of the same charge sign results in a 
nearly background free sample. The remaining backgrounds are due to fake leptons
and their rate, after the requirement of isolated tracks with certain energy
deposits in the hadron calorimeters, is estimated from inclusive jet samples.
No signal is observed in the distributions of relevant 
kinematic variables (see \emph{e.g.} Fig.~\ref{cdf-www}), and also the cross
section limits obtained are compared with predictions for SM and
fermiophobic Higgs boson production in Fig.~\ref{cdf-www}.

\section{\label{charged}CHARGED HIGGS BOSONS}

Charged Higgs bosons have previously been searched for at the LEP Collider, in
their decays to $\tau\bar{\nu}$ or $c\bar{s}$ final states. Assuming that these
two decay modes saturate the $H^{\pm}$ decays, masses below 79 \GeV\ were
excluded.

If $H^{\pm}$ are relatively light, $m(H^{\pm}) < m_{t}-m_{b}$, their dominant
production mechanism at the Tevatron collider is in decays of top quarks,
$t\rightarrow b H^{+}$. This decay mode may be an important one either at low or
at high values of $\tan\beta$ (due to the $m_{t}\cot\beta$ and $m_{b}\tan\beta$
terms in the coupling, respectively). For moderate to high $\tan\beta$ it decays
primarily to $\tau\bar{\nu}$, while for low values the decay is mostly to
$c\bar{s}$.

The D0 Collaboration has re-interpreted existing $t\bar{t}$ production cross
section measurements in the context of a search for the decay
$H^{+}\rightarrow c\bar{s}$ under the pessimistic assumption that $m(H^{\pm})
\approx m_{W}$~\cite{ref:d0-hpm}. The only observable consequence of such a
scenario is an apparent increased branching fraction for hadronic decays of the
$W$ boson; this can be tested by a comparison of the measured di-lepton and
$\ell+$jets $t\bar{t}$ cross sections. From the observed ratio,
$\sigma(\ppbar\rightarrow t\bar{t})_{\ell+\mbox{\small
    jets}}/\sigma(\ppbar\rightarrow t\bar{t})_{\ell\ell} = 1.21^{+0.27}_{-0.26}$,
application of the Feldman-Cousins likelihood ordering
principle~\cite{ref:feldman-cousins} leads to the result
$B(t\rightarrow H^{+} b) < 0.35$ at 95\% CL.

A new CDF analysis, on a 2.2 fb$^{-1}$ dataset, has been carried out looking for
the same decay but under the assumption that $m(H^{\pm}) >
M_{W}$~\cite{ref:cdf-hpm}. The analysis is restricted to $\ell+$jets $t\bar{t}$
candidate events, requiring in addition four jets in the final state, two of
which must be $b$ tagged.

\begin{figure}[htb]
  \centering
  \includegraphics[width=0.4\textwidth]{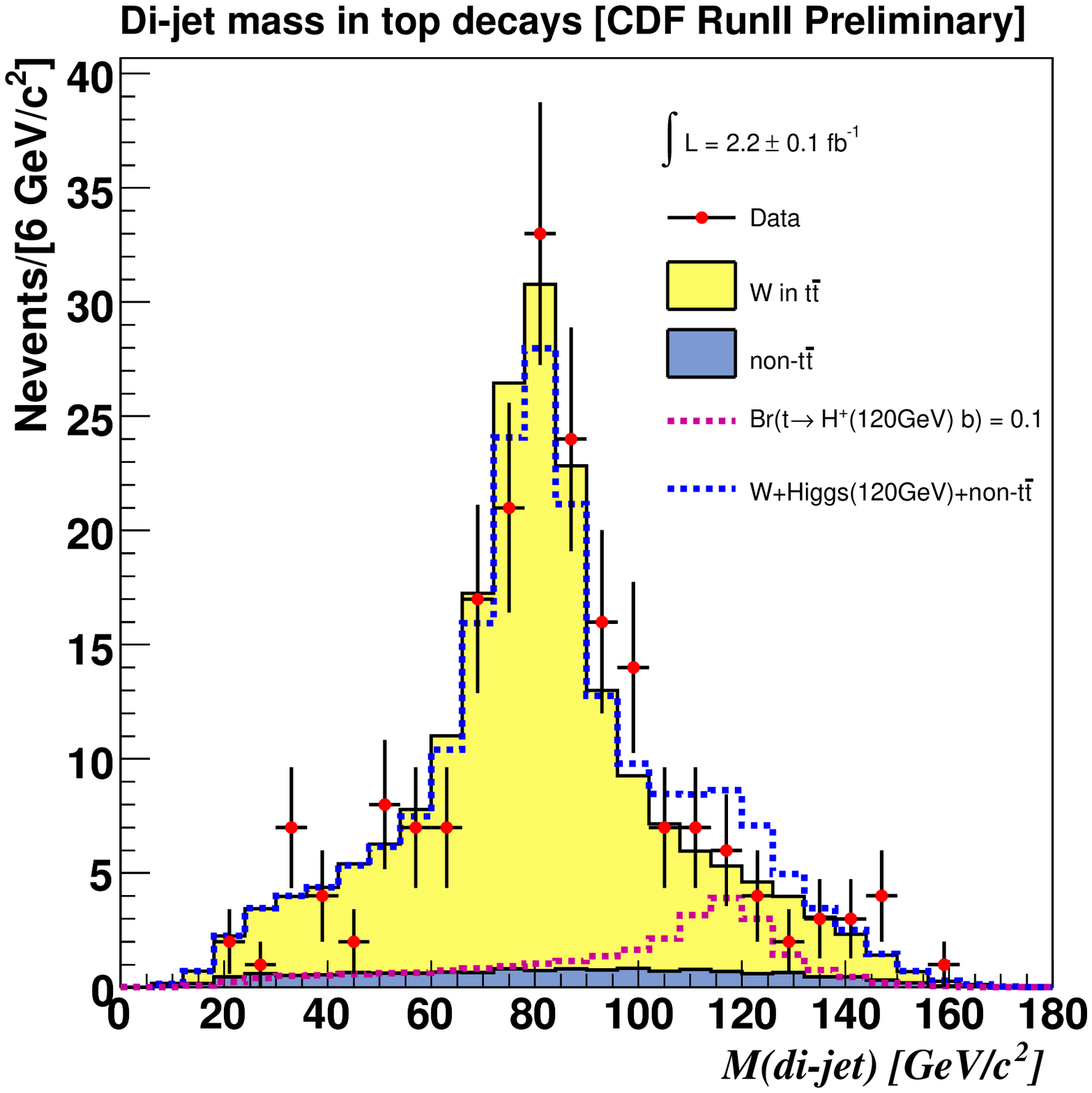}
  \includegraphics[width=0.4\textwidth]{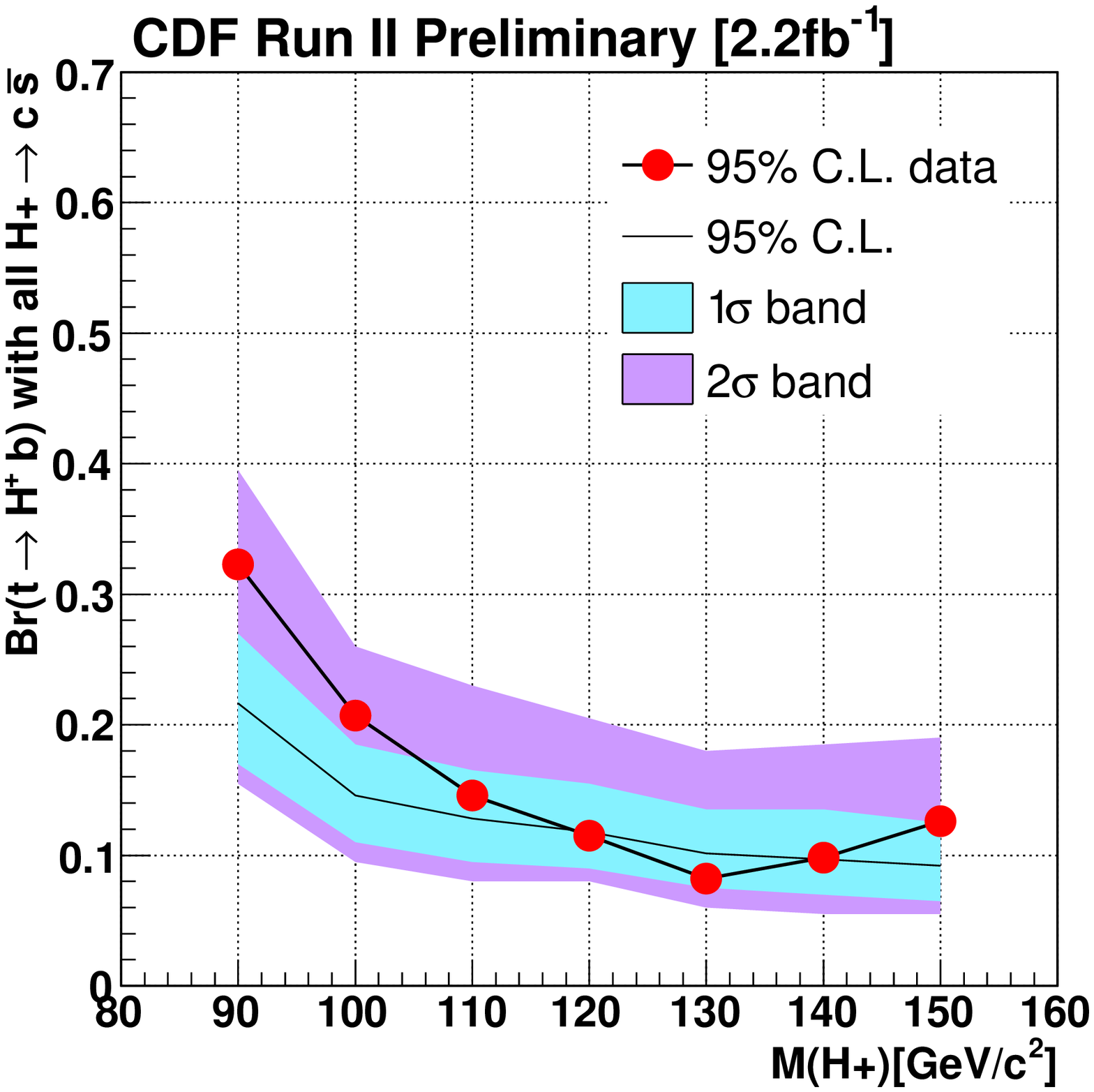}
  \caption{\label{cdf-hpm}Left: di-jet invariant mass in the CDF $t\rightarrow
    H^{+}b$, $H^{+}\rightarrow c\bar{s}$ search. The expected deformation due to
    a 120 \GeV\ $H^{\pm}$ is also shown, for the purpose of illustration. Right:
    resulting upper limits on the product of branching fractions
    $B(t\rightarrow H^{+} b)\cdot B(H^{+}\rightarrow c\bar{s})$.}
\end{figure}

The invariant mass of the remaining two jets is computed. Neglecting backgrounds
and resolution effects, the resulting distribution should yield the W boson
mass. If part of the decays are to charged Higgs bosons, the invariant mass
distribution is modified. A comparison between measured and predicted invariant
mass distributions is shown in Fig.~\ref{cdf-hpm}, with in addition the
expected upper limit at 95\% CL for the signal of a putative charged Higgs boson
of mass 120 \GeV. No such deformation of the di-jet invariant mass distribution
is observed, and upper limits on $B(t\rightarrow H^{+} b)\cdot
B(H^{+}\rightarrow c\bar{s})$ are also given in Fig.~\ref{cdf-hpm}.

\section{\label{doubly}DOUBLY CHARGED HIGGS BOSONS}

A more exotic extension of the SM Higgs sector is to posit the existence of
Higgs multiplets with weak isospin $I > \frac{1}{2}$, leading to doubly charged
Higgs bosons $H^{\pm\pm}$. In general, such extensions are severely constrained
by the $\rho$ parameter, which is $\rho\equiv
m_{W}^{2}/m_{Z}^{2}\cos^{2}\theta_{W} = 1$ at tree level if only Higgs doublets
are present but can be altered significantly in the presence of $I>\frac{1}{2}$
multiplets. However, these constraints can be evaded if the doubly charged Higgs
boson coupling to $W$ bosons vanishes. Two models exist, with ``left-handed''
(``right-handed'') Higgs bosons coupling to left-handed (right-handed) fermions,
respectively.

The D0 Collaboration has searched in a 1.1 fb$^{-1}$ dataset for such doubly
charged Higgs bosons produced through the process
$\ppbar\rightarrow Z/\gamma^{\ast}\rightarrow H^{++}H^{--}$,
with both Higgs bosons decaying to muons,
$H^{\pm\pm}\rightarrow\mu^{\pm}\mu^{\pm}$~\cite{ref:d0-hpp}. As in the case of
the CDF search for fermiophobic Higgs bosons (see Sect.~\ref{fermiophobic}), a
clean sample is obtained by the requirement of two like-sign isolated
muons. Further sensitivity is achieved by requiring a third muon to be present.
The dominant background is from $WZ$ and $ZZ$ events, with muons associated
with a jet passing the isolation criteria. The isolation fake rate is known and
is applied to scale these backgrounds, inferred from simulation. The number of
heavy flavour events (with again muons resulting from semi-muonic decays
appearing to be isolated) is estimated from the number of such events with
opposite charge sign.

\begin{figure}[htb]
  \centering
  \includegraphics[width=0.35\textwidth]{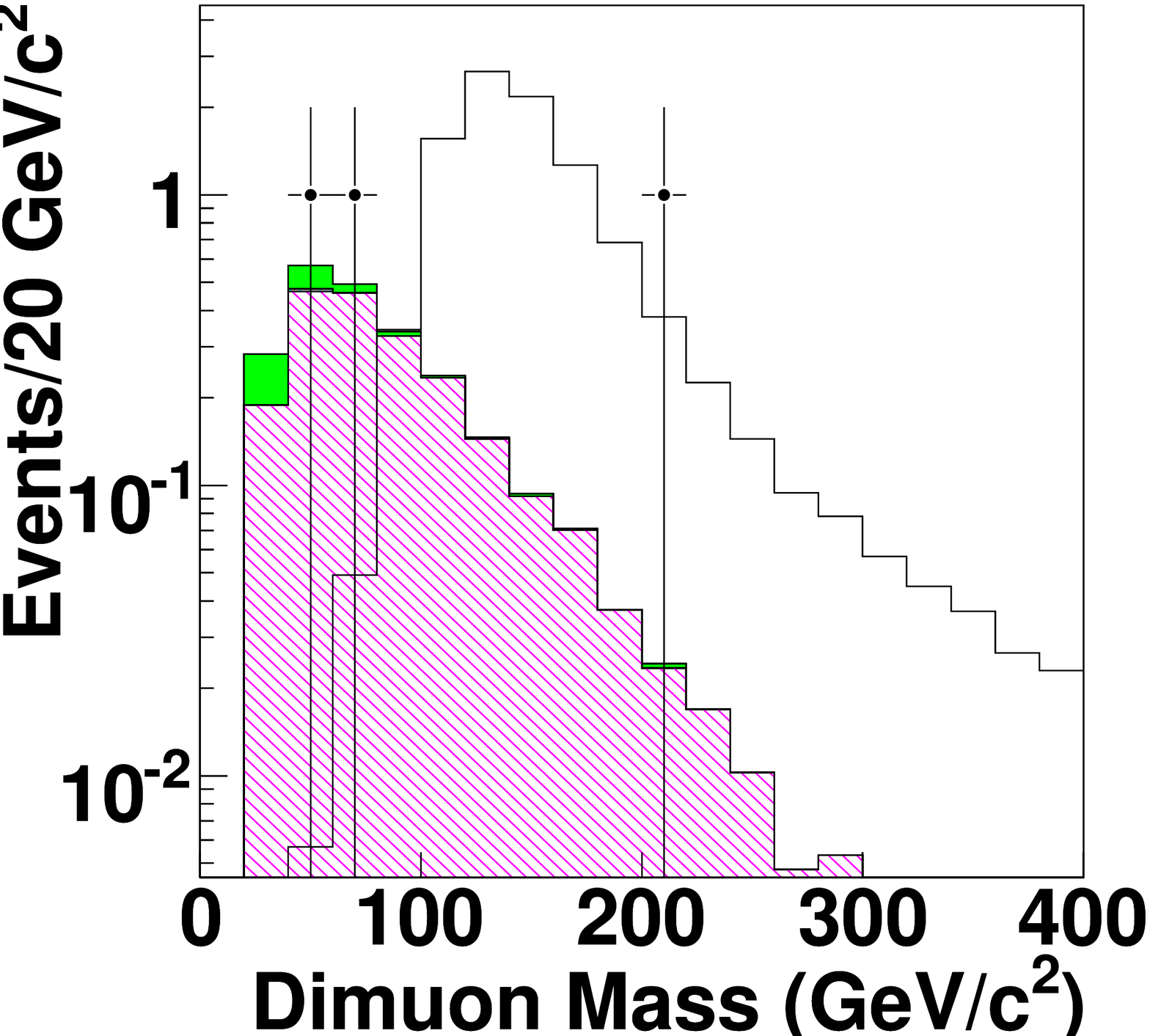}
  \includegraphics[width=0.35\textwidth]{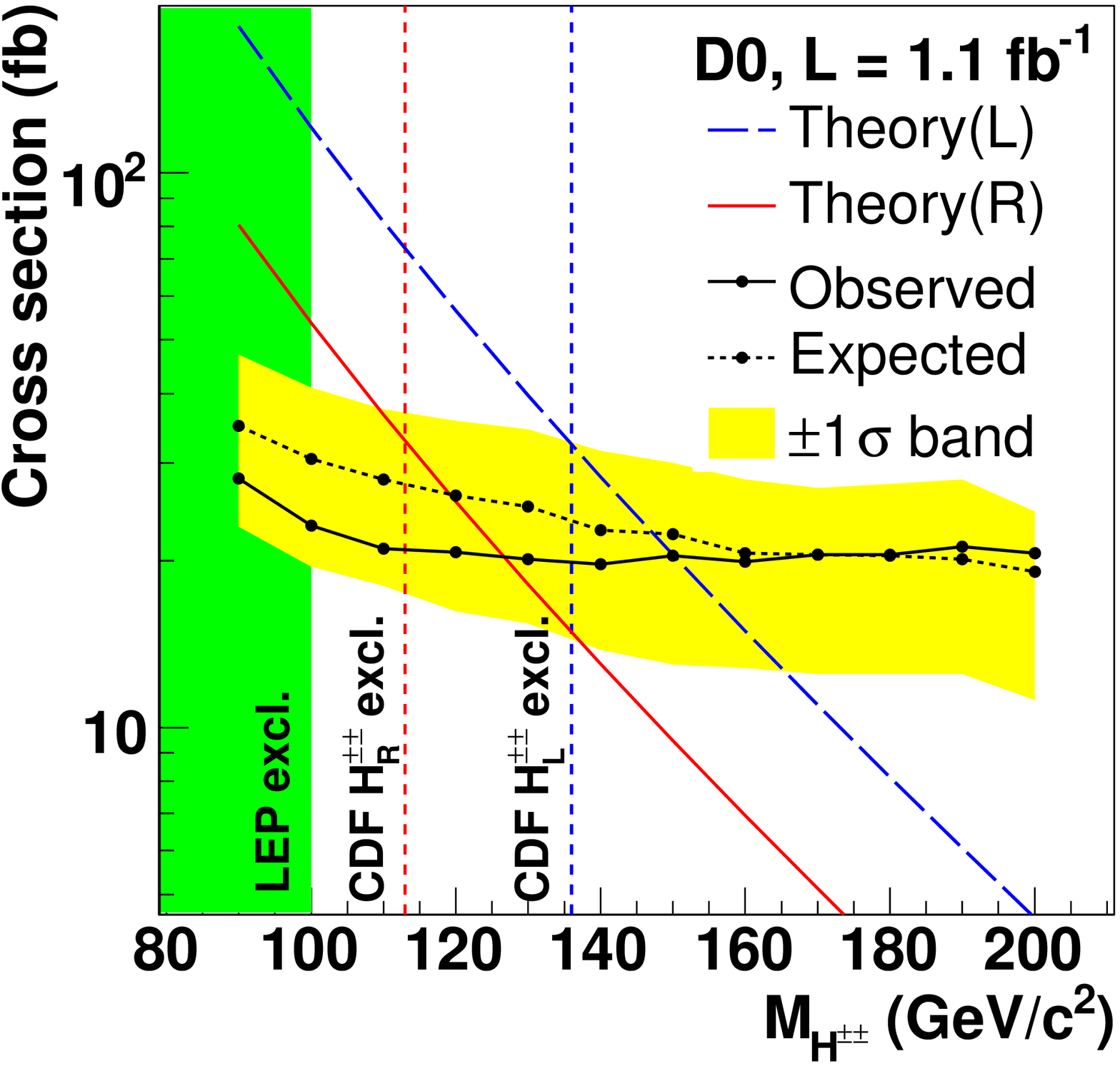}
  \caption{\label{d0-doubly}Left: like-sign di-muon invariant mass after all selection
  criteria in the D0 search for doubly charged Higgs bosons. The dark- and
  light-hatched areas represent the expected multi-jet and $WZ$/$ZZ$
  backgrounds; the open histogram represents the expected signal due to a
  left-handed $H^{\pm\pm}$ with a mass of 140 \GeV. Right: resulting
  cross section limits as a function of the doubly charged Higgs boson mass.}
\end{figure}

Three events remain after all selection criteria, consistent with background
expectations as shown in Fig.~\ref{d0-doubly}. The resulting cross-section
limit, obtained as a function of the Higgs boson mass, is compared with the
theoretical cross-section. This is similarly shown in
Fig.~\ref{d0-doubly}. Lower limits of 150 \GeV (127 \GeV) for left-handed
(right-handed) doubly charged Higgs bosons are obtained.


\end{document}